# Anticipating Formation Trends of Binary and Multiple Asteroids in the Main Asteroid Belt Due to Electric Discharge Effects by Using the Relation of the Sun's Activity and Binary Asteroid's Aggregation Statistics

Mohammad Reza Shafizadeh[1] , Mohammad Reza Shahjahan[2] , Saba hafizi[3]

[1]Astronomical Society of Iran (ASI), Professional Member
shafizadeh@znu.ac.ir
[2]Universal Scientific Education And Research Network (USERN), Iran
[3]Institute of Geophysics, University of Tehran, Iran

**Abstract**. Asteroids are classified as tiny and light objects in the solar system, however some of them possess orbiting moons. According to surveys, 15% of near-earth asteroids have moons. The Electrical Discharge effect provides a new model that describes the formation of single, binary, and triple asteroids influenced by solar magnetic activity. In further studies, collected data was drawn from the aggregation diagram of asteroids based on their average distance from the sun according to various databases, catalogs and spacecraft. The results show that, in areas where solar magnetic activity is high, the binary asteroids are more aggregated. Then, according to results searching catalogs asteroids for shape, spin, and type of orbit, this paper introduces a list of asteroids which, may be transformed from single asteroids to binary (and possibly multiple asteroids) depending on a future increase of solar magnetic activity.

**Keywords.** Asteroid, Binary Asteroid, Electric Discharge, Sun

## Introduction

The electric discharge effect is produced by using an industrial process known as "Electrical Discharge Machining". The machine's EDM mechanism makes excavations, scratches, and cuts, by using electric discharges on any surface from nano scales to macro scales.[1] C.J. Ransom and W. Thornhill introduced this description for planetary objects in 2007 [2]. We introduce the electrical effect as the creator of binary and trinary asteroids because of the metal and stony properties of asteroids and the production of multiple bodies in spatially and temporally concentrated energetic events. We considered the sun as a substantial electric source for shaping the asteroids.[3] we proposed that asteroids can suffer 'cold cathode' discharging in the solar plasma, which can sculpt or cut their surfaces. Nevertheless, the common model of binary asteroids creation is the YORP effect. According to this effect, the sun's light and its photon momentum make the asteroid rotate, and this rotation is the cause of fragmentation over a period of millions of years. One way of testing the electric discharge effect is to obtain the asteroids' concentration statistics in different areas of the Asteroid Belt. Data is grouped according to their semi-major axis, and measurements of the solar electro-magnetic activity in those regions. Therefore, the solar electro-magnetic activity level in different regions and statistical data from binary asteroids aggregation is acquired and the results are charted. According to these charts the binary asteroid's concentration in regions between 1.6-1.9 AU and 2.1-2.3 AU show a remarkable increase. Also the Ulysses spacecraft data illustrate high solar activity in those two regions. [4 , 5]. Therefore this agreement can be considered a confirmation of the electric discharge effect because, based on this theory, as electro-magnetism is stronger, the discharge activity increase and fragmentation of asteroids rise. According to the YORP effect, the binary asteroids should be found in an equal and specified order. Afterwards, the number of Contact Binaries, which, because of their shape, have a greater chance of becoming fragmented, becomes attenuated. Their spin, light curve and mass distribution was examined and a list of Contact Binaries with the highest probability of transforming into Binary Asteroids some catalogs of asteroids in aforesaid regions, according to electric discharge effect was developed.

**Overview on YORP Effect And It's Criticism**

According to this effect, the spin attained by a rotating small object, can be affected by a vast source of light [6] This can change the axis angle of an asteroid which can cause fragmentation over the long term. [7] Among other results of this effect is the changing of rotation poles of an asteroid. [8] This process candeflectthe axis of the asteroidby as much as 90 degrees and, in this way, the axis of the asteroid will reach this state with only light pressure but over a long period of time. It has been noted that this effect is capable of ejecting asteroids from their orbit over a period of long time. [9]

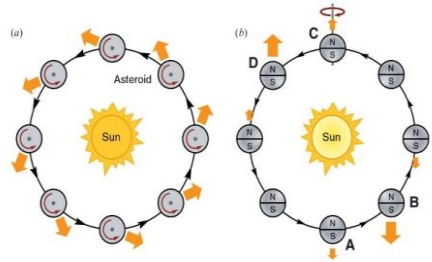

**Figure (1).** YORP Effect explanation

As shown in figure 1, if we suppose the rotating orbit of an asteroid to be a circle and sunlight always radiates on the asteroid's equator, after a period of time the temperature of the equator will rise and thus microwave radiation may reflect from the warm parts of asteroids to space, which has motive force. In this case the asteroid will rotate from warmer parts to colder parts. As shown in figure 1, (a1) after spinning, the asteroid has reached 90 degrees of deflection in it's rotation axis, see figure 1, (b1). In the second figure, after a period time, the north and south poles of the asteroid will receive sunlight that makes the asteroid round again, and this action will repeat over and over. In criticism of this model, the existence of a large number of binary and trinary asteroids with numerous scrapes and incisions on their surface has been mentioned. The photon momentum for producing such shapes in a large variety, requires a long period of time that in some cases would be longer than the age of the solar system. Also no experimental result has yet demonstrated the spin changes by photon collision on stony or metal surfaces. Accepting the valid existence of this mechanism, the number of binary and trinary asteroids should be much fewer than now statistically catalogued and their congestion should be uniformly random around the solar system. However aggregation statistics do not show such results.

**Electric Discharge Effect Explanation**

According to articles about the YORP effect, the angular velocity and the asteroid's spin will change in a billion years, but another impressive parameter was introduced in 2008, The electric discharge effect was proposed by Ransom and Thornhill [2]. This phenomenon can scratch and make incisions on a metallic surface by an electric discharge in a short period of time (in Nano or Micro scales). [1] And this may occur on asteroids. Since 15% of asteroids are binary [3], this may be explained by electromagnetic effects of the Sun and Earth, in fact the metal core of asteroids, beside their remarkable magnetic field, unlike their small body, can be influenced by the Sun, Jupiter or even the

Earth due to their location. When the asteroids move to within the magnetic field of a larger body, the electric discharge effect will occur as a result of charged particle discharge[2, 3] and this event will make scratches and roughen the asteroid's surface.. Whatever the magnetic field that asteroids influence by being stronger, the inequalities will be greater and the probability of fragmentation will be increased. This view is expressed while the number of binary asteroids is much more in comparison with single asteroids [10].Also, it seems unlikely this phenomenon requires billions of years because the asteroids which are near to fragmentation are numerous. We obtained the solar magnetic activity on various locations of The Asteroid Belt from measured data, and compared it with binary asteroid's statistical distribution in terms of semi-major axis (average distance to Sun) data. The results were then drawn as a graph. It shows that the density of asteroids has impressive aggregation on 1.6-1.9 AU and 2.1-2.3 AU, and in addition the data collected by the Ulysses spacecraft indicates that the solar activity in those two ranges shows a steep increase. [4, 5] After that, the number of asteroids named Contact Binaries, which have a greater chance of transforming to Binary Asteroids because of their specific dual core shapes, was checked in the above mentioned ranges, and from 600 asteroids we checked by spins, types, light curves and mass distribution, a list of asteroids with high probability of transforming to Binary Asteroids was produced.

**Binary Asteroid's Density In Asteroid Belt And It's Relation To Solar Activity**

Statistical data of Binary asteroids was taken from the Astronomical Institute of the Academy of Sciences of the Czech Republic [11] and sorted by semi-major axis (average distance to Sun) in MATLAB and by Solar electromagnetic activity. The data was taken from the Ulysses spacecraft.

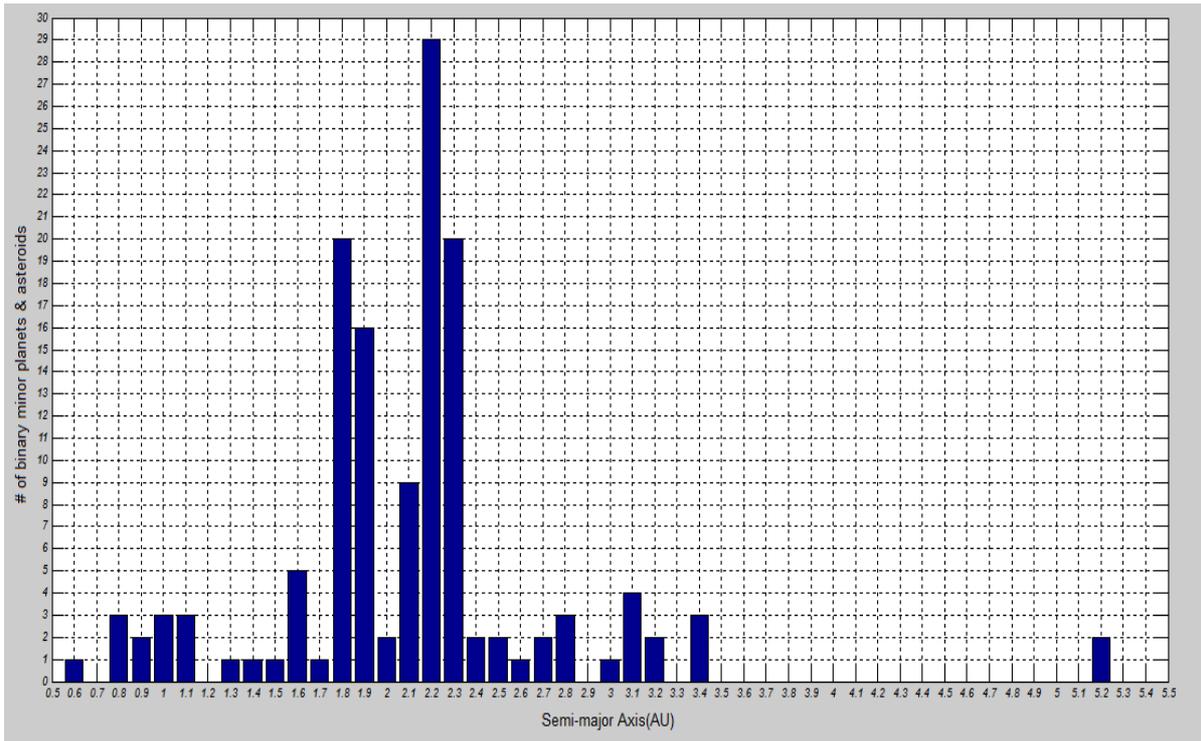

**Graph (1).** Density of Binary asteroids in two distances to sun 1.6-1.9 AU and 2.1-2.3 AU

As shown in Graph(1) , the aggregation of binary asteroids in 1.6-1.9 AU and 2.1-2.3 AU is surprisingly high.

Our assumption and expectation is that the two peaks shown in Graph (1) could be caused by solar spectral changes, and changes in solar activity affecting the Asteroid belt. It could be as a result of the target region moving closer to the sun. We named the closer region to the sun the Z area and the farther region the Y area. Also, the solar activity measure was retrieved from the Ulysses spacecraft as shown in Graph (2)(3)(4). [4,5]

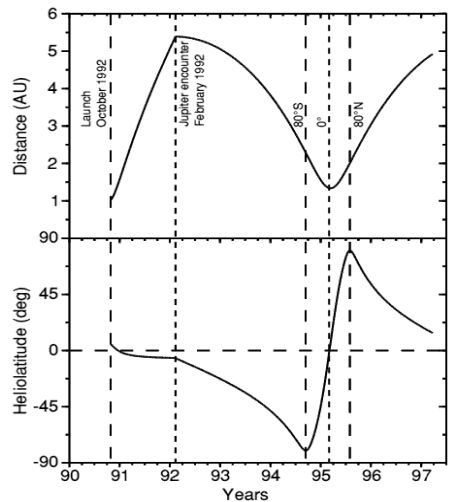

**Graph (2).** The activity of sun in bottom and the distance of spacecraft in top, 1990-1996 [4,5]

As shown in both Graphs, solar activity increases steeply and in 2.4 A.U it reaches the maximum value measured. The separated electric and magnetic data from other spacecraft supports this claim. Therefore it is concluded that these two issues are connected and i indicates that the electric discharge effect is a major cause of Binary Asteroids formation, and, with such a mechanism, formation of Binary Asteroids in The Solar System can be possible.[17]  Graph (3) ,(4) are showing that the separated electric and magnetic data of the sun measured at different distances is another proof of the electric discharge effect as a Binary Asteroids creator.[4,5]

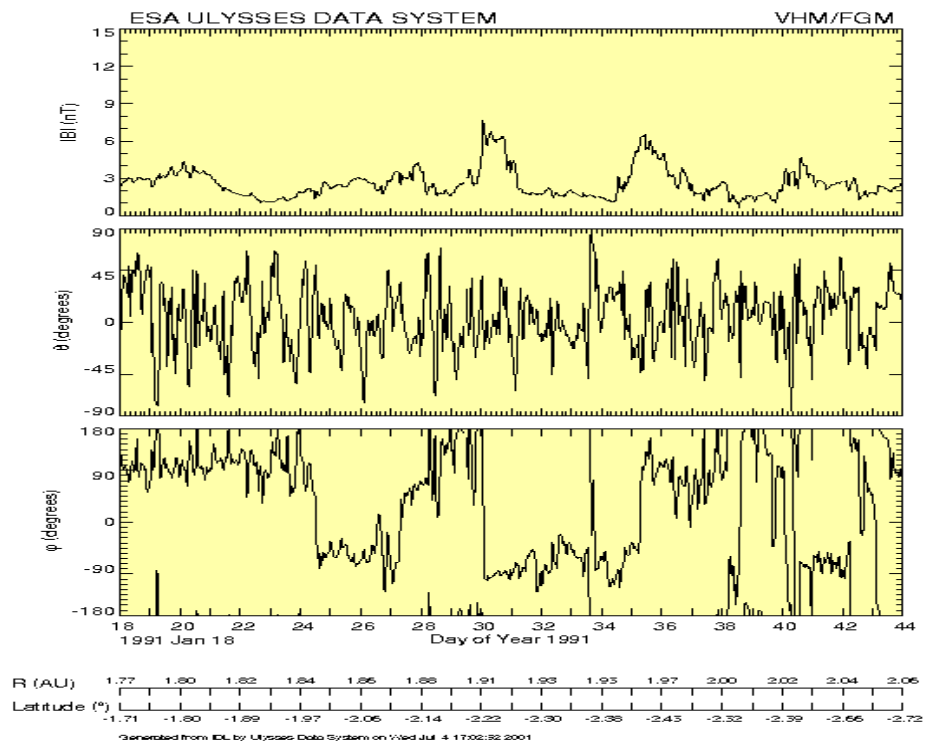

**Graph (3).** Electric and Magnetic data from sun in different distances which includes the Z area
[4,5]

| Asteroid name | Apogee (A.U) | Perigee (A.U) | Semi-major axis (A.U) |
|---|---|---|---|
| 433 Eros | 1.7 | 1.1 | 1.4 |
| 4769 Castalia | 1.6 | 0.55 | 1.06 |
| 2013 JB36 | 2.15 | 1.06 | 1.61 |
| 3169 Ostro | 2.01 | 1.76 | 1.89 |
| 1999 CF9 | 2.83 | 0.70 | 1.77 |
| 2002 TD66 | 2.85 | 0.86 | 1.85 |
| 2005 NZ6 | 3.42 | 0.24 | 1.83 |
| 2005 CR37 | 2.79 | 1.01 | 1.90 |
| 1620 Geographos | 1.6 | 0.8 | 1.2 |
| 2004 EW9 | 2.94 | 0.89 | 1.91 |

**Table (1).** Anticipated Asteroids in Z area

## List Of Anticipating The Formation Trend Of Binary And Multiple Asteroid For Single Asteroids In Z and Y Areas

In the final step of this work, a list of some Asteroids, which according to mentioned assumptions can be transformed to binary or multiple asteroids in the future, is introduced by observing the appearance, spin and light curves of those single asteroids that pass through the Z and Y areas. In this list, the apogee, perigee, and semi-major axis parameters is included. It is apparent how much asteroids in the Z and Y regions through their orbital motion have an increased chance of fragmentation, thus the chance of becoming a binary or multiple asteroid is higher. For achieving this list, databases such as Astrosurf, Space Frieger, DAMIT, Johnston list , JPL small body database and articles which are published for some of them were used and from Asteroids' catalogs these cases were chosen.[16-12] It should be mentioned that these shapes and appearances can be found in other regions of The Asteroid belt.According to our first reviews they are aggregated in close proximity to Jupiter and it may be that this aggregation is caused by Jovian electromagnetic activity in those areas, but this hypothesis needs more research and study.

| Asteroid name | Apogee (A.U) | Perigee (A.U) | Semi-major axis (A.U) |
|---|---|---|---|
| 4179 Toutaris | 4.12 | 0.93 | 2.53 |
| 6489 Golevka | 4 | 1 | 2.5 |
| 4486 Mithra | 3.6 | 2.7 | 2.2 |
| 2013 JR28 | 3.64 | 0.76 | 2.20 |
| 951 Gaspra | 2.59 | 1.82 | 2.20 |
| 5535 Annefrank | 2.35 | 2.07 | 2.21 |
| 9969 Braille | 3.35 | 1.32 | 2.34 |
| 44 Nysa | 2.78 | 2.06 | 2.42 |
| 43 Ariadne | 2.57 | 1.83 | 2.20 |
| 4179 Toutaris | 4.12 | 0.93 | 2.53 |

**Table (2).** Anticipated Asteroids in Y area

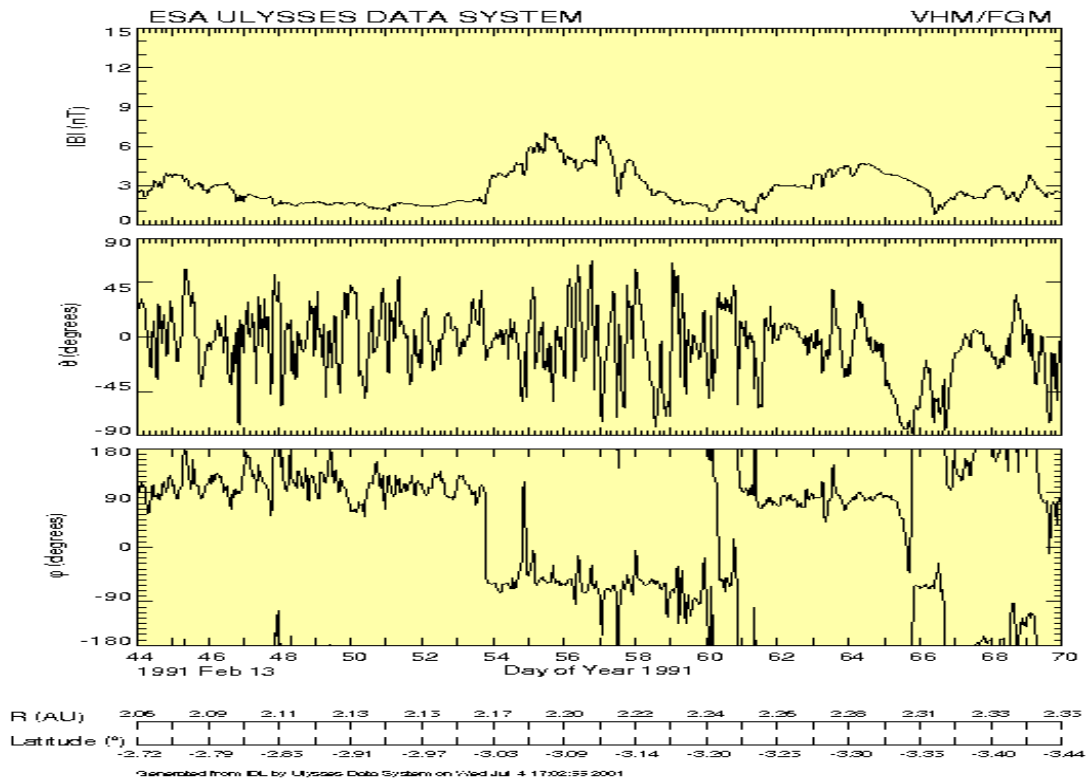

**Graph (4).** Electric and Magnetic data from sun in different distances which includes the Y area [4,5]

**Overview of Electric Discharge Effect In Some Asteroids** In this statistical data study, there was an interesting point: the aggregation of Asteroids with two mass centers (Contact binaries) are exceptionally high in the Z and Y areas. The data for three Asteroids published in other articles are presented below as example of electric discharge effect on binary making:

**4486 Mithra Asteroid**

Much good information and accurate data has been collected for this Asteroid by both the Doppler Effect and visual observation [18]. The method consists of accurately measuring the reflected waveform from an Asteroid and then simulating it's shape by wavelength changes. The mass distribution can be obtained by this method as well.

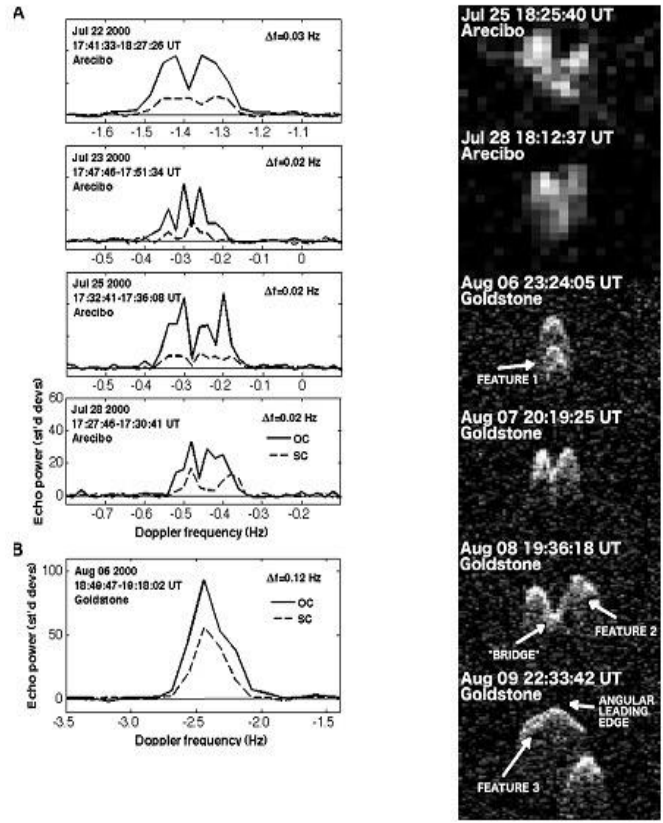

**Figure (2).** Radio data collected using the Doppler Effect from 4486 Mithra Asteroid [18]

As shown in Figure (2), the asteroid is transformed to a two-center object, called a Contact Binary. The middle part of the asteroid is sunken and two heads can be seen on both sides of it. These kinds of asteroids are greater in number precisely in the regions in which binary asteroids are numerous. In Figure (3). The mass distribution of asteroid 4486 Mithra is shown.

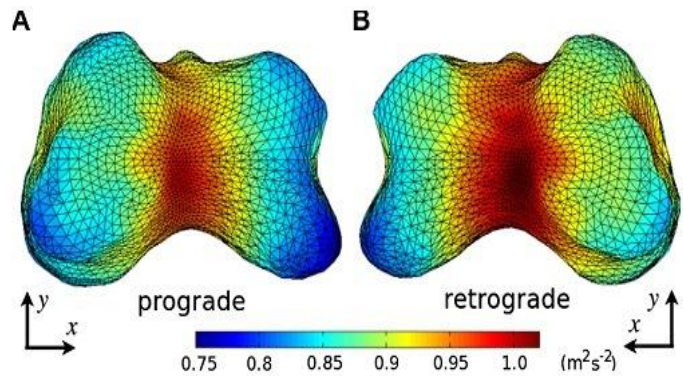

**Figure (3).** Mass distribution of 4486 Mithra, red parts shows low mass and blue parts shows high mass distribution.

The center part of this asteroid is reduced in diameter because of an electrical discharge between it and the sun. This process continues to reshape the asteroid and may eventually divide it into two separate objects that revolve around each other. It may be presumed that the number of electrical discharges will be in proportion to the metal density in the asteroid and its distribution on the asteroid's body. This asteroid may well be transformed into a binary asteroid in the future with increasing solar activity.

**44 Nysa and 43 Ariadne Asteroids**

According to observational data and light curves drawn for these asteroids, [19] they are binary system in which the moon is attached to the main body, and thus they can be considered as Binary Asteroids like 4486 Mithra but have a better chance to become separated. In addition to these asteroids, all objects on the introduced list have similar properties and their aggregation and distribution in discussed regions are remarkable. (Fig.4)

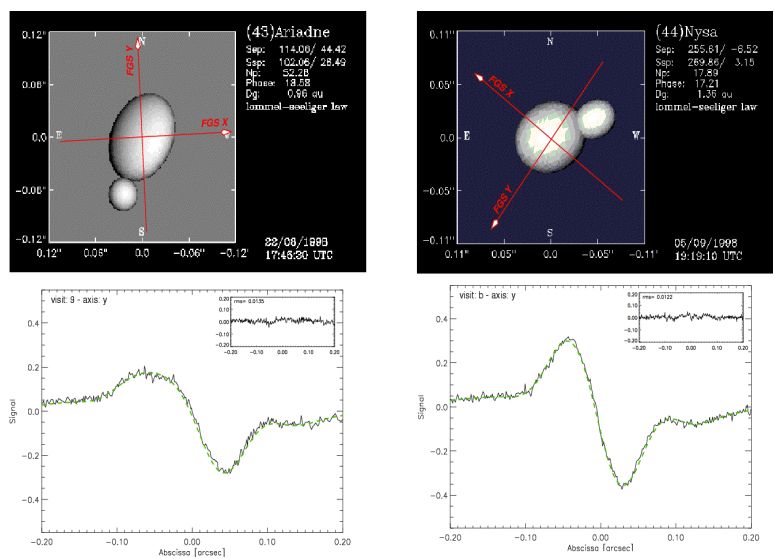

**Figure (4).** 44 Nysa in right and 43 Ariadne in left, Asteroids shapes are simulated by their light curves [19]

The light curves drawn for these two Asteroids indicates that they have a strange stretched and dual core appearance that may be considered to be the result of discharges in a linear manner on the Asteroid's surface.

**Conclusion**

The Binary and Multiple Asteroids distribution in the Asteroid Belt is not random or disarrayed. As the electric discharge effect in Binary Asteroids formation showed, the important and main cause for this is an electrical discharge which originated from a powerful electromagnetic source, the sun. This assertion can be demonstrated by a statistical study of distribution and aggregation of asteroids and measurement of electromagnetic solar activity at different distances. According to this survey and data collected from different databases, the Solar Electromagnetic activity level and the aggregation of Binary Asteroids have a direct correlation, and this can explain the high number of Binary Asteroids in two regions in the asteroid belt at 1.6-1.9 A.U and 2.1-2.3 A.U. Relying on this assumption and after checking more than 600 Asteroids, it is suggested that some of them have a greater chance to split and shrink into smaller components revolving around each other based on their distance from sun.

## Acknowledgment

We wish to thank Ms. Annis Scott , Prof. Don.Scott and Mr. Wal Thornhill for their support and help.